\documentclass[prl,twocolumn,amsmath,superscriptaddress,showpacs]{revtex4}

\usepackage{graphicx}
\usepackage{amssymb}
\usepackage{amsmath}
\usepackage{epsfig}

\begin{document}

\title{Emergence of zero-lag synchronization in generic mutually coupled chaotic systems}

\author{Meital Zigzag}
\affiliation{Department of Physics, Bar-Ilan University, Ramat-Gan,
52900 Israel}
\author{Maria Butkovski}
\affiliation{Department of Physics, Bar-Ilan University, Ramat-Gan,
52900 Israel}
\author{Anja Englert}
\affiliation{Institute for Theoretical Physics, University of
W\"urzburg, Am Hubland, 97074 W\"urzburg, Germany}
\author{Wolfgang Kinzel}
\affiliation{Institute for Theoretical Physics, University of
W\"urzburg, Am Hubland, 97074 W\"urzburg, Germany}
\author{Ido Kanter}
\affiliation{Department of Physics, Bar-Ilan University, Ramat-Gan,
52900 Israel}

\begin{abstract}
Zero-lag synchronization (ZLS) is achieved in a very restricted
mutually coupled chaotic systems, where the delays of the
self-coupling  and the mutual coupling are identical or fulfil some
restricted ratios. Using a set of multiple self-feedbacks we
demonstrate both analytically and numerically that ZLS is achieved
for a wide range of mutual delays. It indicates that ZLS can be
achieved without the knowledge of the mutual distance between the
communicating partners and has an important implication in the
possible use of ZLS in communications networks as well as in the
understanding of the emergence of such synchronization in the
neuronal activities.
\end{abstract}


\maketitle

Two identical chaotic systems starting from almost identical initial
states, end in completely uncorrelated trajectories\cite{1,1a}. On
the other hand, chaotic systems which are mutually coupled by some
of their internal variables often synchronize to a collective
dynamical behavior\cite{2,3}. The emergence of synchronization plays
important functioning roles in natural and artificial coupled
systems. One of the most fascinating collective dynamical behavior
is the zero-lag synchronization (ZLS), known also as an isochronal
synchronization. ZLS or nearly ZLS was measured  in the activity of
the brain between widely separated cortical regions\cite{4a,5a,6a},
where synchronization of neural activity has been shown to underlie
cognitive acts\cite{3a}. The mechanism of the ZLS phenomenon has
been subject of controversial debate, where the main puzzle is how
two or more distant dynamical elements can synchronize at zero-lag
even in the presence of non-negligible delays in the transfer of
information between them.

The phenomenon of ZLS was also experimentally observed in the
synchronization of two mutually chaotic semiconductor lasers, where
the optical path between the lasers is a few orders of magnitude
greater than the coherence length of the
lasers\cite{einat1,phase_sync,ingo1,shutter}, and the analogy
between the spiking optical pattern and the neuronal spiking was
also recently established\cite{spiking}. This phenomenon has
attracted a lot of attention, mainly because of its potential for
secure communication over a public channel\cite{einat1}. In
\cite{hilbert} it was recently shown that it is possible to use the
ZLS phenomenon of two mutually coupled symmetric chaotic systems for
a novel key-exchange protocol generated over a public-channel.  Note
that in contrary to a public scheme which is based on mutual
coupling, private-key secure communication is based on a
unidirectional coupling\cite{Roy,nature} and it is susceptible to an
attacker which has identical parameters and is coupled to the
transmitted signal. The generation of secure communication over a
public channel requires mutual coupling and was only proven to be
secure based on the ZLS phenomenon\cite{hilbert}.

Recently, it has been shown both numerically and analytically that
various architectures of coupled chaotic maps can exhibit ZLS\cite{
7a,sub_lat1}. The main disadvantage of this phenomenon is that ZLS
even between two mutually coupled chaotic systems can be achieved
only for very restricted architectures and it is highly sensitive
for mismatch between the delays of the mutual coupling and the
self-feedback. These delays have to be identical or have to fulfil
special ratios. Such a realization might exist in a time-independent
point-to-point communication, but it is far from the realm of
communications networks.


In this letter we first demonstrate the constraint that ZLS is
achieved only for very restricted ratios between the self-feedback
and the mutual delays, $n\tau_d=m\tau_c$, where $n$ and $m$ are
(small) integers. We next show that one can overcome this constraint
when multiple self-feedbacks are used. For the simplicity of the
presentation we mainly concentrate on the Bernoulli map, where
results of simulations can be compared to an analytical
solution\cite{physica_d,sub_lat1}. However we observed the reported
phenomenon for other chaotic maps and systems as well, and it is
exemplified by the ZLS of mutually couple chaotic semiconductor
lasers, depicted by the Lang-Kobayashi differential
equations\cite{LK,einat1}.

\begin{figure}[h]
\includegraphics[scale=0.6]{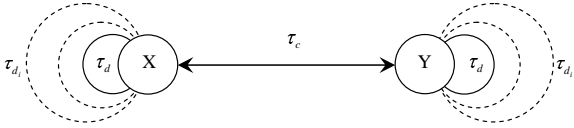}
\caption{\label{schem} A schematic diagram of two mutually coupled
units at a distance $\tau_c$ with one self-feedback with a delay
equals to $\tau_d$ (solid lines). Additional self-feedbacks are
denoted by the dashed lines.}
\end{figure}

The cornerstone of our system is the simplest chaotic map, the
Bernoulli map, $f(x)=(a x)mod1$, which is chaotic for $a
> 1$. The dynamical equations of the two mutually coupled chaotic
units, $X$ and $Y$, with one self-feedback (see solid lines in
figure \ref{schem}) are given by
\begin{equation}
\label{dymanical}
\begin{split}
\!\!\!\!x_{t}=(1-\varepsilon)f(x_{t-1})+\varepsilon[\kappa
f(x_{t-\tau_{d}})+(1-\kappa)f(y_{t-\tau_{c}})]\\
\!\!\!\!y_{t}=(1-\varepsilon)f(y_{t-1})+\varepsilon[\kappa
f(y_{t-\tau_{d}})+(1-\kappa)f(y_{t-\tau{c}})]
\end{split}
\end{equation}
where $\tau_d$ and $\tau_c$ are the delays of the self feedback and
the mutual coupling, respectively\cite{sub_lat1}. The quantities
$1-\varepsilon$, $\varepsilon\kappa$ and $(1-\kappa)\varepsilon$
stand for the strength of the internal dynamics, self-feedback and
the mutual coupling, respectively.

The stationary solution of the relative distance between the
trajectories of the two mutually coupled chaotic Bernoulli maps can
be analytically examined\cite{physica_d,sub_lat1}. Let us denote by
$\delta x_{t}$ and $\delta y_{t}$ small perturbations from the
trajectories $x_{t}$ and $y_{t}$, respectively. Using the ansatz
$\delta x_{t}=c^t\delta x_{0}$ and $\delta y_{t}=c^t\delta y_{0}$
and linearizing equations (\ref{dymanical}), one can find the
characteristic polynomial
\begin{equation}
\label{polynomial}
\!\!\!\!c-a(1-\varepsilon)-a\varepsilon\kappa c^{ 1-\tau_{d} }+a\varepsilon(1-\kappa)c^{1-\tau_{c}}=0
\end{equation}
where $\lambda=ln|c|$ is the Lyapunov exponent. Simulations of the
dynamical equations (\ref{dymanical}) and the semi-analytical
calculation of the maximal Lyapunov exponent of the characteristic
polynomial (\ref{polynomial}), indicate that ZLS is the stationary
solution of the dynamics   only when the delays of the self-feedback
and the mutual coupling fulfil the constraint
\begin{equation}
\label{delta1} n\tau_{d}+m\tau_{c}=0
\end{equation}
where the available integers for  $n, m\in \mathbb{Z}$ are functions
of $\epsilon$ and $\kappa$. Results are exemplified in figure 2 for
$\epsilon=0.9$ and $\kappa=0.8$ (left panel) and for $\epsilon=0.9$
and $\kappa=0.4$ (right panel). For the left panel, ZLS is achieved
for the pairs $(m,n)=(-1,n)$ where $n=1,2, ..., 10$ and
$(3,-1)$\cite{comment2}. For the right panel ZLS is achieved for the
pairs $(-1,n)~ n=1, ..., 4$, $(3,-1)$, $(5,-1)$, $(7,-1)$, $(3,-2$)
and $(5,-2)$\cite{even}. Those lines may have width, so the more
accurate equation is $|n\tau_d-m\tau_c|\le\delta$, where
$\delta\thickapprox 2$.
\begin{figure}[h]
\includegraphics[scale=0.50]{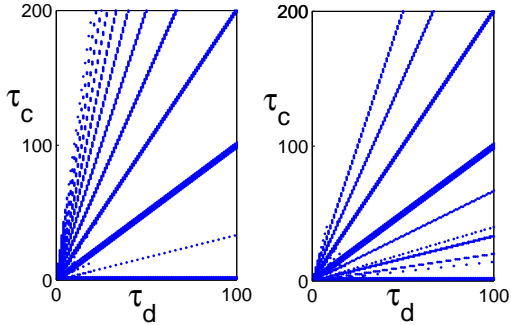}
\caption{\label{tau_d tau_c} Simulations and semi-analytic results
for the ZLS points in the phase space ($\tau_d, \tau_c$) with
$a=1.1, \varepsilon=0.9$ and $\kappa=0.8$ left panel and
$\varepsilon=0.9$ and $\kappa=0.4$ right panel.}
\end{figure}
The constraint  (\ref{delta1}) indicates that ZLS can be achieved
only when $\tau_c$ is accurately known, which is far from the realm
of communications networks. In order to increase the possible ZLS
range of $\tau_c$ for a fixed $\tau_d$, we added more
self-feedbacks, as depicted in figure \ref{schem}. The generalized
dynamical equations for the case of multiple self-feedbacks are
given by
\begin{eqnarray}
x_{t}\!\!=\!\!(1-\varepsilon)f(x_{t-1})\!\!+\!\!\varepsilon[\kappa\sum_{l=1}^{N}\alpha_{l}
f(x_{t-\tau_{d_l}})\!\!+\!\!(1-\kappa)f(y_{t-\tau_{c}})]\nonumber\\
y_{t}\!\!=\!\!(1-\varepsilon)f(y_{t-1})\!\!+\!\!\varepsilon[\kappa\sum_{l=1}^{N}\alpha_{l}
f(y_{t-\tau_{d_l}})\!\!+\!\!(1-\kappa)f(x_{t-\tau_{c}})]
\end{eqnarray}
where $N$ stands for the number of self-feedbacks and the parameter
$\alpha_{l}$ indicates the weight of the $l^{th}$ self-feedback
fulfilling the constraint $\sum_{l=1}^{N}\alpha_{l}=1$. In order to
reveal the interplay between possible $\tau_c$ which lead to ZLS and
a given set of $\{\tau_{d_l}\}$ we first examine in detail the case
of $N=2$.

\begin{figure}[h]
\includegraphics[scale=0.37]{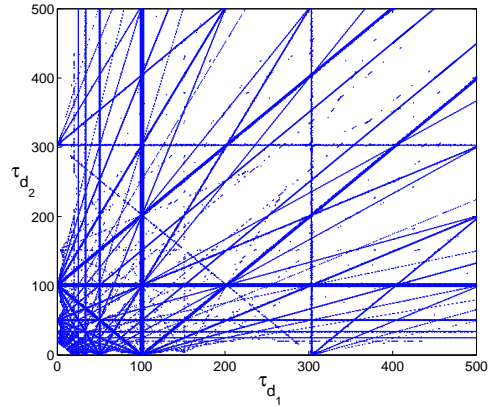}
\caption{\label{tau_d1 tau_d2} Simulations and semi-analytic results
for the ZLS points in the phase space ($\tau_{d_1}, \tau_{d_2}$) for
$\tau_c=101,~a=1.1, \varepsilon=0.9$, $\kappa=0.8$ and
$\alpha_i=1/2$.}
\end{figure}

Results of simulations with $N=2$ which were confirmed by the
calculation of the largest Lyapunov exponent obtained from the
solution of the characteristic polynomial, similar to equation (2),
are depicted in  figure \ref{tau_d1 tau_d2}. The synchronization
points $(\tau_{d_1},~\tau_{d_2})$ where ZLS is achieved form
straight lines. A careful analysis of the equations of these lines
indicates that their equations are
\begin{equation}
\label{delta2}
n_{1}\tau_{d_1}+n_{2}\tau_{d_2}+m\tau_{c}=0
\end{equation}
where $n_{1}, n_{2}$ and $m$ are integers. The lines may have a
small width, hence a more accurate equation for the ZLS points is
$|n_{1}\tau_{d_1}+n_{2}\tau_{d_2}+m\tau_{c}|\le\delta$, where
$\delta\thickapprox 2$. The same equations for the ZLS lines and
with similar possible width, $\delta$, were obtained in simulations
with different $\epsilon,~\kappa,~\alpha_i$ and  $\tau_{c}$, prime
and non-prime numbers in the range $[31, 720]$.

\begin{figure}[h]
\includegraphics[scale=0.5]{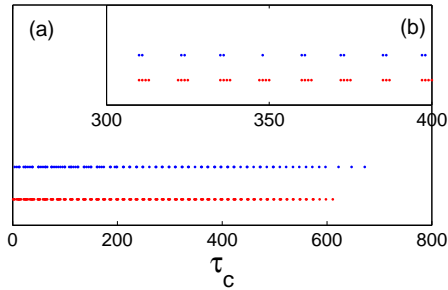}
\caption{\label{comparison} ZLS for two mutually coupled Bernoulli
maps with $a=1.1, \varepsilon=0.9$ and $\kappa=0.8$ and with two
self-feedbacks,$\tau_{d_1}=25, \tau_{d_2}=87$. ZLS points obtained
from the exhaustive search of eq. (5) with $m=\pm1,\pm3$ and $n_i$
in the range $[-6,6]$ (blue points). The ZLS points obtained from
simulation and semi-analytical results (red points). The inset is a
blow up of a section of possible $\tau_c$ with ZLS.}
\end{figure}

Figure 3 indicates that for $\epsilon=0.9$ and
$\kappa=0.8$\cite{comment3} , for instance,  $m$ can take the
integers $\pm1$ and $\pm3$ only. In order to examine the possible
range of the integers $\{n_{i}\}$ we ran an exhaustive search
simulation, $-6 \leq n_{i}\leq 6$ and $m=\pm1,~\pm3$, and obtained
integer $\tau_{c}$ from equation (\ref{delta2}). Figure
\ref{comparison} depicts results of such an exhaustive search and
the analytical solution of appropriate characteristic polynomials.
The comparison between the results indicates the following two main
conclusions: (a) $n_i\thickapprox6$ gives a similar synchronization
range,
(b) the lines have an extension of up to $2$, hence the actual ZLS
points fulfil the equation
$|n_{1}\tau_{d_1}+n_{2}\tau_{d_2}+m\tau_{c}|\le2$ (see the inset of
figure \ref{comparison}).
Note that a few blue points are missing in the ZLS obtained in the
semi-analytical solution (red points) indicating that a few
combinations $(n_1, n_2, m)$ are missing.
We also analyze in detail the case of
triple self-feedbacks, equation (4) with $N=3$, and find that  ZLS
points follow the equation
$|n_{1}\tau_{d_1}+n_{2}\tau_{d_2}+n_{3}\tau_{d_3}+m\tau_{c}|\le\delta\sim2$,
and in this case the ZLS points form planes.

The generalization of the ZLS points for $N=1, 2$ and $3$ to the
case of multiple self-feedbacks is
\begin{equation}
\label{multi_delta}
\sum_{i=1}^{N} n_{i}\tau_{d_i}+m\tau_{c}=0
\end{equation}
where $n_i$ and $m$ take bounded integer values.
This generalization was indeed confirmed in
simulations and solving the characteristic polynomials with up to
$N=7$.

In order to obtain a continuous range of $\tau_{c}$ for which ZLS is
achieved, we examined the scenario of $4$ different
$\tau_{d_i}=11,15,18,150$. We select one remarkably large $\tau_d$
such that we can see its effect on the range of $\tau_{c}$ where ZLS
is achieved. To measure the quality of the ZLS we used the
correlation function, which is defined by
\begin{equation}
\label{correlation}
C=\frac{\langle x_{t}y_{t}\rangle-\langle x_{t}\rangle\langle y_{t}
\rangle}{\sqrt{\langle x^2_{t}\rangle-{\langle x_{t}\rangle}^2}\sqrt{\langle y^2_{t}\rangle-{\langle y_{t}\rangle}^2}}
\end{equation}
where $C=1$ indicates complete ZLS and $\langle ... \rangle$ stands
for an average over the last $1000$ time steps.  The correlation
function, $C$, obtained in simulations is depicted in figure 5 and
indicates the following results. Multiple self-feedbacks result in a
continuous range of ZLS for $\tau_c$, hence it is not required to
know exactly the mutual distance (value), $\tau_c$.
\begin{figure}[h]
\includegraphics[scale=0.5]{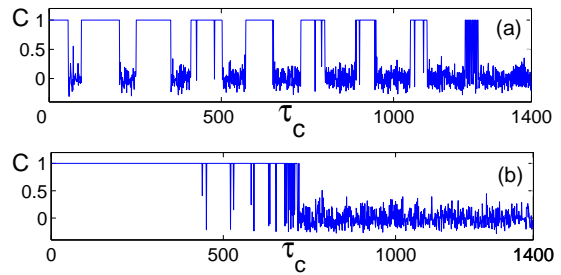}
\caption{\label{multiple} Simulations results of the correlation,
$C$,  as a function of $\tau_c$ for  $a=1.1, \varepsilon=0.9$ and
$\kappa=0.8$ and  four $\tau_{d_i}=11,15,18,150$. The weight of the
self-feedbacks, $\alpha_i$ in eq. (4), are  in (a)
$\alpha_1=\alpha_2=\alpha_3=0.25/3$ and $\alpha_4=0.75$ and in (b)
$\alpha_1=\alpha_2=\alpha_3=0.65/3$ and $\alpha_4=0.35$ }
\end{figure}
Panel (a) of figure 5 indicates that there are at least $7$
continuous  ZLS regimes, each one of them is centered at $150n_4$,
where $n_4=0,1,~...,~6$ and the plateaus are extended $\sim\pm60$
around the centers (slightly decreases with increasing $n_4$). This
width, $\pm60$, is much smaller than the ZLS range of the only three
short self-feedbacks $11,15,18$ which was found to be $\sim[1,150]$,
indicating that the effective $n_1,n_2$ and $n_3$ in eq. (6) are
less than $6$. This discrepancy is a result of the dominated weight
of $\tau_4=150$, $\alpha_4=0.75$, in figure 5(a). For a smaller
weight for the largest delay $150$, $\alpha_4=0.35$, panel (b) of
figure 5, a ZLS is continuously achieved up to $\tau_c\sim700$. In
this case a weak weight for the largest delay results in limited
$n_4$ which takes the values of $0,1,2,3,4$ only, and we expect ZLS
in four continuous regimes centered around $\tau_c=0,~150,~300,~450$
and $600$\cite{Anja}. However these four regimes are now merged  by
the $\pm150$ width inspired by the strengthened weight for the short
self-feedbacks, $\alpha_1=\alpha_2=\alpha_3=0.65/3$\cite{comment1}.

In the general case there is an interplay between the following
three parameters characterizing the set of the delay times:
$\tau_{d_{max}}$ which is comparable to $\tau_{c}$,
$\{\tau_{d_{i}}\}\ll\tau_{d_{max}}$ and
$\Delta_{i}=\tau_{d_{i+1}}-\tau_{d_{i}}$ $i=1,...~N-2$, where
$\{\tau_{d_{i}}\}$ are arranged in an increasing rank order. For
instance, the following three sets of four self-feedbacks
$(2,6,9,150),~(11,15,18,150),~(80,84,87,150)$ are characterized by
the same $\tau_{d_{max}}$, $\Delta_1$ and $\Delta_2$. What is the
main difference between the ZLS profile of these sets and which set
maximizes the continuous range of ZLS? The first set opens only a
small continuous ZLS regime ($\sim20$ for parameters of panel (a))
around $150n_4$, since the time delays are very short. The third set
almost does not open a continuous regime of ZLS, since
$\tau_{d_1},\tau_{d_2},\tau_{d_3} \gg \Delta_1,\Delta_2$. The
maximal continuous ZLS range is achieved when  short delays
$\tau_{d_1},\tau_{d_2},\tau_{d_3}$ are comparable with
$\sim6\Delta_1,6\Delta_2$ (see eq. (6)) which is a case of the
second set.

Most of the reported simulations were carried out from close initial
conditions, however, one can find $(\epsilon,\kappa)$ such that ZLS
is achieved from random initial conditions at a comparable time to
ZLS with only one time delay, $\tau_c\!=\!\tau_d$.

Similar results were obtained also for mutually coupled chaotic
logistic maps where the Lyapunov exponent is fluctuating in time and
is positive only on the average.

Finally we report that a similar phenomenon of ZLS occurs in
simulations of two mutually coupled semiconductor lasers depicted by
the Lang-Kobayashi equations\cite{LK}. Our simulations are based on
the version and the parameters of these equations as in
\cite{einat1}, with additional time delays. Figure 6(a) depicts the
ZLS as a function of $\tau_c$ for the case of $4$ time delays
$\tau_{d_i}=3,4,5,20 ns$ and $\kappa_i=\sigma=30 ns^{-1}$ and in
figure 6(b) for $6$ time delays $\tau_{d_i}=11,12,13,14,15,16 ns$
with $\kappa_i=\sigma=25 ns^{-1}$, where for both cases the
threshold current was  $p=1.02$. For each $\tau_c$ the duration of
the simulation was $7000 ns$ and the emergence of ZLS was estimated
from the measured cross correlation of the last $20$ windows of $100
ns$\cite{comment}. Results indicate that for the case of $6$ delays
ZLS is achieved in the range $\sim[1,80]ns$ where for the case of
$4$ time delays for $\sim[1,45]ns$\cite{shore}. These
synchronization regimes can be explained by equation (6) with
$n_i=0,\pm1,\pm2$ only. It is consistent with our simulations of
only one time delay where ZLS is achieved for $\tau_d=n\tau_c$ with
$n=1,2,3$ only (instead of $1, ...,~6$ for the examined maps) . Note
that no extension on a time scale of $ns$ is expected, $\delta=0$,
however, preliminary results indicate that a similar phenomenon
occur where $\Delta_i=0.01ns$ which is comparable with the coherence
length of the laser.

The research of I.K. is partially supported by the Israel Science
Foundation.

\vspace{-0.3cm}
\begin{figure}[h]
\includegraphics[scale=0.55]{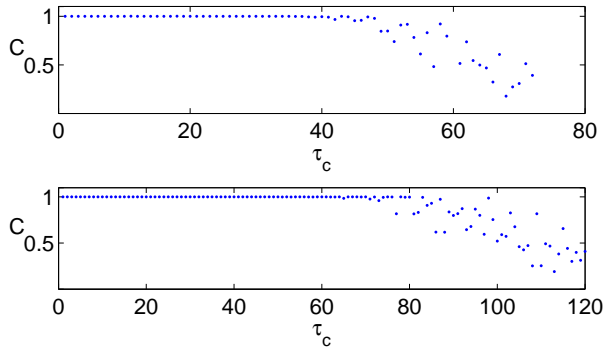}
\caption{\label{multiple} Simulation results of the correlation, eq.
(7), for two mutually coupled semiconductor lasers (details in the
text).
Panel (a) for $4$ delays $\tau_d=3,4,5,20ns$ and
panel (b) for $6$ time delays ~$\tau_d,=11,12,13,14,15,16ns$.}
\end{figure}


\begin{thebibliography}{99}
\bibitem{comment2} For $\tau_c\sim1$, ZLS is achieved independent of $\tau_d$
due to the interernal dynamics, see eq. (1).
\bibitem{even} Even integers $m$ appear, for instance, for $\epsilon=0.2$,
$\kappa=0.6$ where ZLS is achieved for $(2,-3)$ and $(2,-5)$. The
range of $(m,n)$ also increase in the limit of weak chaos
$a\rightarrow 1^+$.
\bibitem{comment3} Note that the effective weight of the mutual coupling is $\alpha\kappa=0.4$
but the weight of the self-feedback is $1-\kappa=0.2$. Hence, the
ZLS is not expected to be equal to either panels of figure 2.
\bibitem{Anja} One can find analytically an upper bound for $n_i^{max}(\alpha_i, \epsilon, \kappa)$ which
for $\epsilon=0.9$ and $\kappa=0.8$, for instance,  is
$n_i<18.8\alpha_i$ ,which is consistent with our results.
\bibitem{comment1} The semi-analytical solution indicates that plateaus of ZLS are
slightly wider and no few sudden drops among the plateaus. These
tiny mismatches are due to almost zero maximal Lyapunov exponent
close to the plateaus boundaries, and limited number of steps in
simulations.
\bibitem{comment} For threshold current $p=1.02$, chaotic signals consnist of low frequency
fluctuations (LFFs) where  short desynchronizations occur with the
used numerical integration $10^{-13}s$, see also \cite{einat1}. We
avoid this affect by calculating the average cross correlation (eq.
7) of the maximal $10$ among last $20$ windows of size $100ns$ each.
\end{thebibliography}
\end{document}